# Resonant Raman Scattering and Optical Absorption Studies of Zn(II) Impurities in L-Alanine Single Crystal

A. Nonato[1,*], G.G.S. Teles[1], C. C. Silva[1], R. X. Silva[2], Juan S. Rodríguez-Hernández[3], C.W.A. Paschoal[3], B. A. V. Lima[4], A. S. de Menezes[5], C.C. Santos[5,*]

[1]Coordenação de Ciências Naturais - Física, Universidade Federal do Maranhão, Centro de Ciências de Bacabal - CCBA, 65700-000, Bacabal - MA, Brazil

[2]Centro de Ciência e Tecnologia em Energia e Sustentabilidade, Universidade Federal do Recôncavo da Bahia, 44085-132, Feira de Santana - BA, Brazil

[3]Departamento de Física, Universidade Federal do Ceará, Campus do Pici, 65455-900, Fortaleza - CE, Brazil.

[4]Coordenação de Ciências Naturais - Química, Universidade Federal do Maranhão, Centro de Ciências de São Bernardo - CCSB, 65550-000, São Bernardo - MA, Brazil

[5]Departamento de Física, Universidade Federal do Maranhão, Centro de Ciências de Exatas – CCET, 65080-805, São Luís - MA, Brazil

[*]Corresponding author. Tel: +55 (98) 98205-4803

E-mail address: ariel.nonato@ufma.br

E-mail address: clenilton.cs@ufma.br

# ABSTRACT

We present a study of low-concentration Zn (II) impurities in L-Alanine Single Crystal (ZNLA) employing resonant Raman scattering (RRS) and optical absorption spectroscopy. By analyzing the normalized integrated intensity of selective vibrational modes, we observe a resonant enhancement of some specific Raman-active modes near the UV absorption band, which are associated either with lattice vibrations or with intramolecular modes, the latter being related to methyl and carboxyl groups in ZNLA. These findings indicate that zinc cations occupy interstitial positions interacting with amino and carboxyl groups of L-alanine, as previously reported for $Fe^{3+}$- and $Cu^{2+}$- doped L-alanine crystals. Based on the optical absorption measurements, the direct bandgap was determined to be approximately 3.8 eV. Overall, our approach proves to be useful for detecting interactions between transition metal ion impurities and intramolecular moieties in transition metal ion doped amino acid systems.

**Keywords:** L-alanine, Transition metals, Resonant Raman scattering, X-ray diffraction.

# 1. Introduction

Nonlinear optical single crystals have recently attracted significant interest due to their various physical phenomena and properties, which make them suitable for plenty of applications such as lasers, optical computing, optical communication, optoelectronics, optical switching, optical modulation, optical information storage devices, optical limiting, and signal transmission [1–10]. Organic nonlinear optical (NLO) materials have been regarded as a strong alternative in research fields due to their intriguing characteristics, including high thresholds for laser damage and optical propagation, broad third-order nonlinear optical sensitivity and coefficients, as well as strong chemical stability [3,11]. Particularly, amino acid crystals have been standing out for their efficiency in second harmonic generation (SHG) [12–14]. Furthermore, it has recently been shown that single crystals of the amino acid L-threonine could be used as optical waveguides and filters within the telecommunications spectrum [15].

L-alanine has been recognized as a unique organic NLO material within the amino acid family, featuring an orthorhombic crystal structure with space group $P2_12_12_1$ ($D_2^4$). It is well known that the physical properties of L-alanine crystals strongly depend on their structure. Numerous studies have investigated the growth, morphology, optical, thermal, mechanical, and electrical properties of L-alanine doped with different metal ions [14,16–22]. However, few studies have investigated the effects of dopants on their vibrational properties [23–25]. Most studies primarily focus on the thermal stability of L-alanine under extreme temperature and pressure conditions [26–28]. In fact, at low dopant concentrations, these amino acid-based structures generally do not form complexes with metal ions. Instead, the dopant tends to occupy interstitial positions, usually not inducing significant structural changes [29–31].

In this context, resonant Raman scattering (RRS) has been proven to be a powerful technique for investigating the interactions between interstitial impurities and specific molecular groups, which may involve intramolecular (localized) or lattice (extended) vibrations, within the L-alanine crystal lattice [32]. Tuning the laser excitation wavelength can lead to an enhancement of specific vibrational modes in the Raman spectra, especially when resonance conditions are achieved. The phonons that exhibit such enhancement are those that couple directly with the electronic transitions involved in the resonance process. This technique has been successfully employed to investigate the impact of Cu(II) and Fe(III) impurities on the vibrational properties of L-alanine single crystals by varying the excitation line from 514.5 nm to 454.5 nm [24]. This approach revealed a resonant enhancement of some of the Raman-active lattice vibrations, including displacements of the hydrogen bonds between the carboxyl and amino groups of neighboring molecules [23,24].

In this paper, we investigated the effects of interstitial zinc impurities on the optical, structural, and vibrational properties of L-alanine single crystals. For this purpose, pure L-alanine (LA) and zinc-doped L-alanine (ZNLA) single crystals were synthesized using the slow evaporation method. A comprehensive Raman scattering analysis was performed using fixed laser excitation lines (405, 532, 633, and 785 nm) to explore the interactions of zinc ions with intramolecular groups in the L-alanine structure. Additionally, optical absorption measurements were conducted to understand the resonant and antiresonant behavior exhibited by specific Raman-active modes in ZNLA crystal. To the best of our knowledge, the study of the crystal structure revealing the position of $Zn^{2+}$ impurities and their interactions with intramolecular vibrations in L-alanine single crystals has not yet been reported. Finally, our results are compared with previously reported data on Fe- and Cu-doped L-alanine single crystals. Based on these findings, we discuss the interaction of zinc impurities in the L-alanine crystal lattice.

## 2. Experimental procedures

### *2.1 Material, synthesis and crystal growth*

The crystals used in this study were grown using the slow evaporation method at room temperature. Pure L-alanine single crystals were also synthesized using the same technique [33]. For the Zn-doped L-alanine crystals (ZNLA), 0.2716 g of L-alanine was dissolved in 10 mL of distilled water. To this solution, 0.1740 g of zinc nitrate was added, and the mixture was stirred magnetically for approximately one hour. After this period, small amounts of sodium hydroxide were gradually added to the solution until the pH was adjusted to 8. The resulting solution was then filtered and transferred to a beaker, covered with plastic film with small holes in order to allow slow water evaporation, and kept at room temperature. The photographs of the as-grown pure and doped crystals are shown in **Figure 1**.

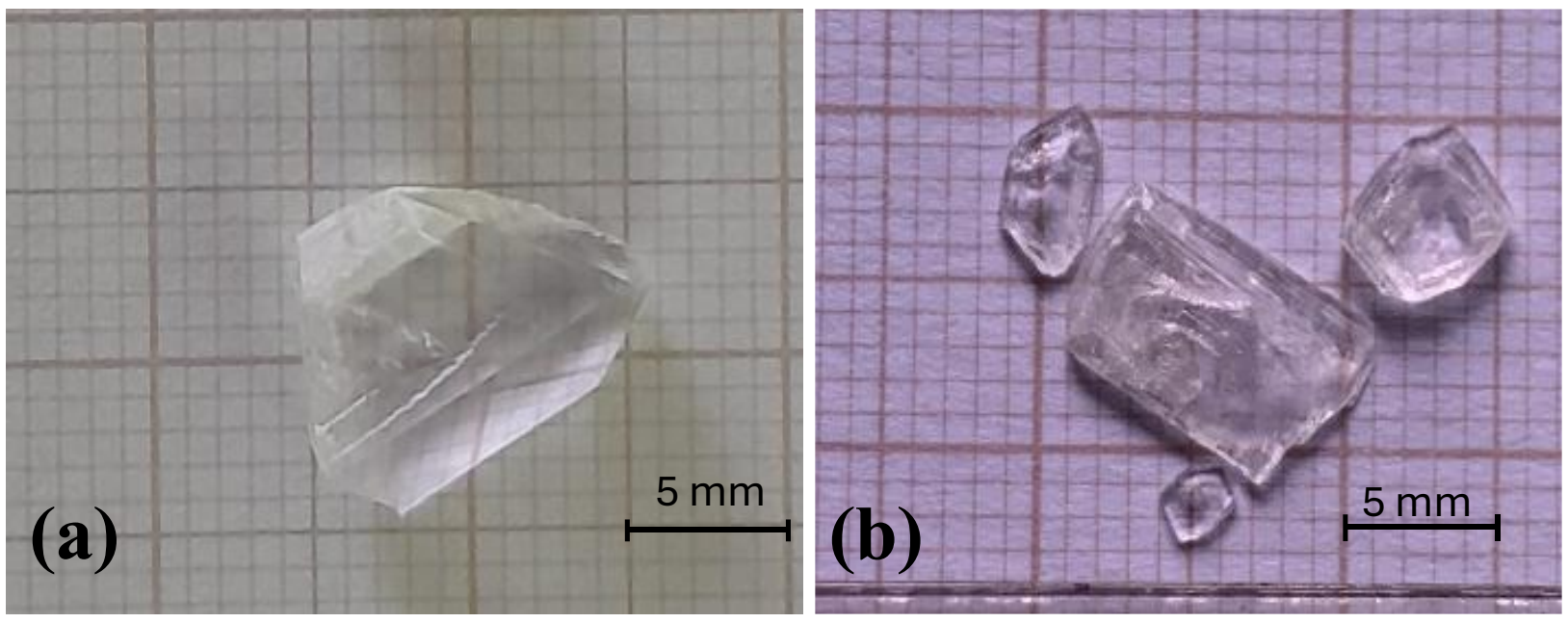

**Figure 1.** Photographs of the as-grown (a) pure L-alanine crystal (LA) and (b) Zn-doped L-alanine single crystal (ZNLA).

### *2.2 Characterization methods*

An X-ray diffractometer (XRD) (Bruker, model D8 Advance) equipped with Cu Kα radiation was used to assess the crystal structure and phases within the 2θ range of 5 to 60°. The X-ray diffraction measurements were performed on the powdered samples of LA and ZNLA, and the resulting diffractograms were refined by the Rietveld method using the GSAS suite [34]. The Raman spectra data were acquired using a T64000 micro-Raman spectrometer from Horiba/Jobin-Yvon, equipped with a nitrogen-cooled charge-coupled-device (CCD) detector. The single crystals were excited with four different laser source lines at 405, 532, 633, and 785 nm, with the laser power adjusted to prevent heat damage. The light was focused on the sample using an Olympus BX41 microscope with an SLMPLN 20× objective lens (WD = 26.5 mm). The laser was focused, and depending on the wavelength, the circular spot size varied between 2.0 and 4.0 μm in diameter, with a power of 3 mW. The Raman signal was dispersed with an 1800 gr/mm grating. The Raman spectra were fitted with Lorentzian functions using the software Fityk 0.9 [35]. All Raman spectra were normalized using the integrated intensity of the 264 $cm^{-1}$ band as a reference. This specific vibrational mode was selected due to its relatively stable intensity across all measured spectra, exhibiting the lowest variation when compared to other spectral features. Room-temperature absorption spectra were recorded through diffuse reflectance spectroscopy using a Shimadzu UV-2600 spectrophotometer equipped with an ISR-2600 Plus integrating sphere, covering the wavelength range of 250–600 nm. The reflectance spectrum obtained was further processed by the Kubelka-Munk function to transform it into an absorption spectrum.

# 3. Results and discussions

The PXRD analysis confirmed that both crystals exhibit an orthorhombic structure with four molecules ($C_3H_7NO_3$) per unit cell (Z = 4), belonging to $P2_12_12_1$ ($D_2^4$) space group [36]. The powder X-ray diffraction (PXRD) patterns of pure LA and ZNLA samples are shown in **Figure S1 (a-b)**, and lattice parameters obtained for the pure LA and ZNLA crystals are presented in **Table S1**. A comparison reveals differences in the cell parameters between LA and ZNLA. These changes in the crystal lattice parameters can be attributed to the incorporation of zinc nitrate into the L-alanine structure, resulting in the expansion of the unit cell volume without significant changes to the crystal structure. Based on the lattice parameter data from **Table S1**, and assuming approximately the same coordination for all cations, **Figure 2** presents the dependence of the L-alanine unit cell volume on the ionic radius of different dopants. The expansion or contraction of the crystal lattice unit cell strongly depends on the ionic radius of the dopant, as has been observed in other doped-based L-alanine systems.

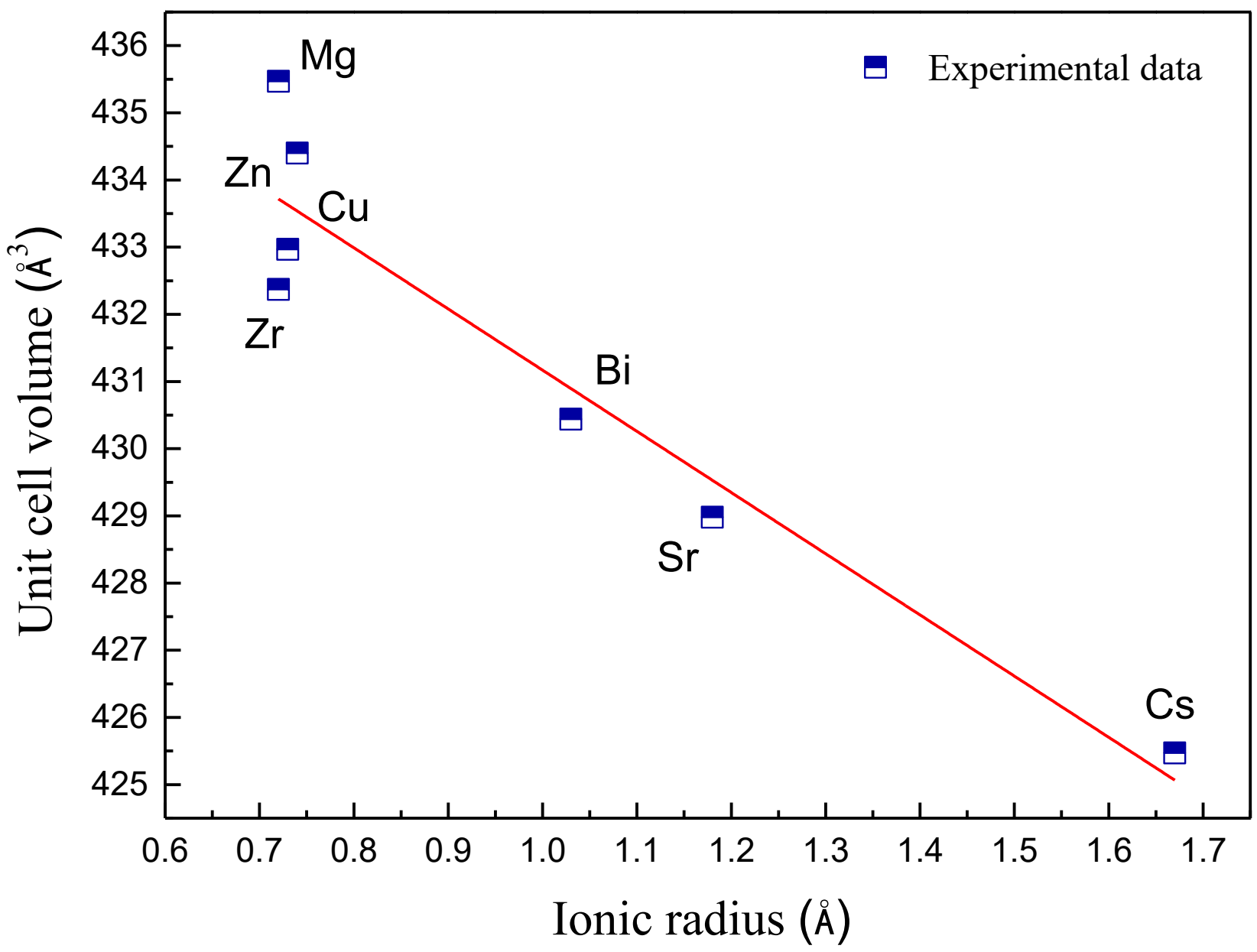

**Figure 2.** Dependence of the unit cell volume of L-alanine on the ionic radius of different dopants. All ionic radio presented here has been taken from "Revised Effective Ionic Radii and Systematic Studies of Interatomic Distances in Halides and Chalcogenides" by R. D. Shannon [37]. The red solid line represents a guide for the eyes.

According to the symmetries and the Wyckoff site occupations in the LA crystalline unit cell within the $P2_12_12_1$ ($D_2^4$) space group [38], group theoretical analysis predicts 153 optical modes, which are distributed into irreducible representations of the $D_2$ factor group as $39A \oplus 38B_1 \oplus 38B_2 \oplus 38B_3$ [39]. All these vibrational modes are Raman-active, while modes with B symmetry representations are also IR-active. A summary of the factor group analysis of the fundamental vibrations of L-alanine single crystal is presented in Table 1.

Table 1. Factor group analysis of fundamental vibrations of L-Alanine crystal.

| Atoms | Wyckoff Sites | Site Symmetry | Irreducible representation | | | |
|---|---|---|---|---|---|---|
| | | | A | $B_1$ | $B_2$ | $B_3$ |
| C1 | 4a | $C_1$ | 3 | 3 | 3 | 3 |
| C2 | 4a | $C_1$ | 3 | 3 | 3 | 3 |
| C3 | 4a | $C_1$ | 3 | 3 | 3 | 3 |
| H1 | 4a | $C_1$ | 3 | 3 | 3 | 3 |
| H2 | 4a | $C_1$ | 3 | 3 | 3 | 3 |
| H3 | 4a | $C_1$ | 3 | 3 | 3 | 3 |
| H4 | 4a | $C_1$ | 3 | 3 | 3 | 3 |
| H5 | 4a | $C_1$ | 3 | 3 | 3 | 3 |
| H6 | 4a | $C_1$ | 3 | 3 | 3 | 3 |
| H7 | 4a | $C_1$ | 3 | 3 | 3 | 3 |
| N | 4a | $C_1$ | 3 | 3 | 3 | 3 |
| O1 | 4a | $C_1$ | 3 | 3 | 3 | 3 |
| O2 | 4a | $C_1$ | 3 | 3 | 3 | 3 |
| $\Gamma_{Total} = 39A \oplus 39B_1 \oplus 39B_2 \oplus 39B_3$ \| $\Gamma_{Acoustic} = B_1 \oplus B_2 \oplus B_3$ | | | | | | |
| $\Gamma_{Raman} = 39A \oplus 38B_1 \oplus 38B_2 \oplus 38B_3$ \| $\Gamma_{IR} = 39B_1 \oplus 39B_2 \oplus 39B_3$ | | | | | | |

In **Figure 3**, we present the room temperature Raman spectra of Zn-doped L-alanine single crystal excited by different laser lines, ranging from 405 nm to 785 nm. For comparison, the Raman spectra performed on LA and ZNLA powder samples are also shown in **Figure 3**. The spectra of the powder samples of LA and ZNLA are quite similar, showing no changes in their profile due to the presence of zinc impurities, even within the 2500–3500 $cm^{-1}$ spectral region (not shown in the figure). This observation is consistent with previous findings, where no significant spectral changes were detected for Zn-complexed L-alanine [40]. The spectra of the ZNLA single crystal were collected by focusing the laser on its flat surface at the same position (shown in the inset of **Figure 3**). The ZNLA single crystal exhibits a preferential orientation, which explains why some modes present in the ZNLA single crystal do not appear or are not well defined in the powder ZNLA and LA spectra (see **Figure S2** in the supplementary material). Also, this spectrum resembles those obtained for L-alanine single crystal with signal orientation $a(cc)b$ according to the reference [41]. The general features of the polarized Raman spectra of L-alanine have been extensively discussed in references [41–43]. By changing the excitation line from 785 nm down to 405 nm, significant variations in the spectral intensities of specific modes are observed, while others remain unchanged. One of the vibrational modes exhibiting enhanced intensity corresponds to the lattice mode at 115 $cm^{-1}$, which has its origin in the intermolecular interactions involving hydrogen bonds between the carboxyl and amino groups of neighboring molecules (N-H···O) [44–46]. On the other hand, the modes appearing in the region between 1300 and 1550 $cm^{-1}$ exhibit a reduction in their intensities (as indicated in **Figure 3b**). All observed Raman-active modes are listed in **Table S2** along with their assignment to lattice and intramolecular vibrations, which will be discussed in detail later.

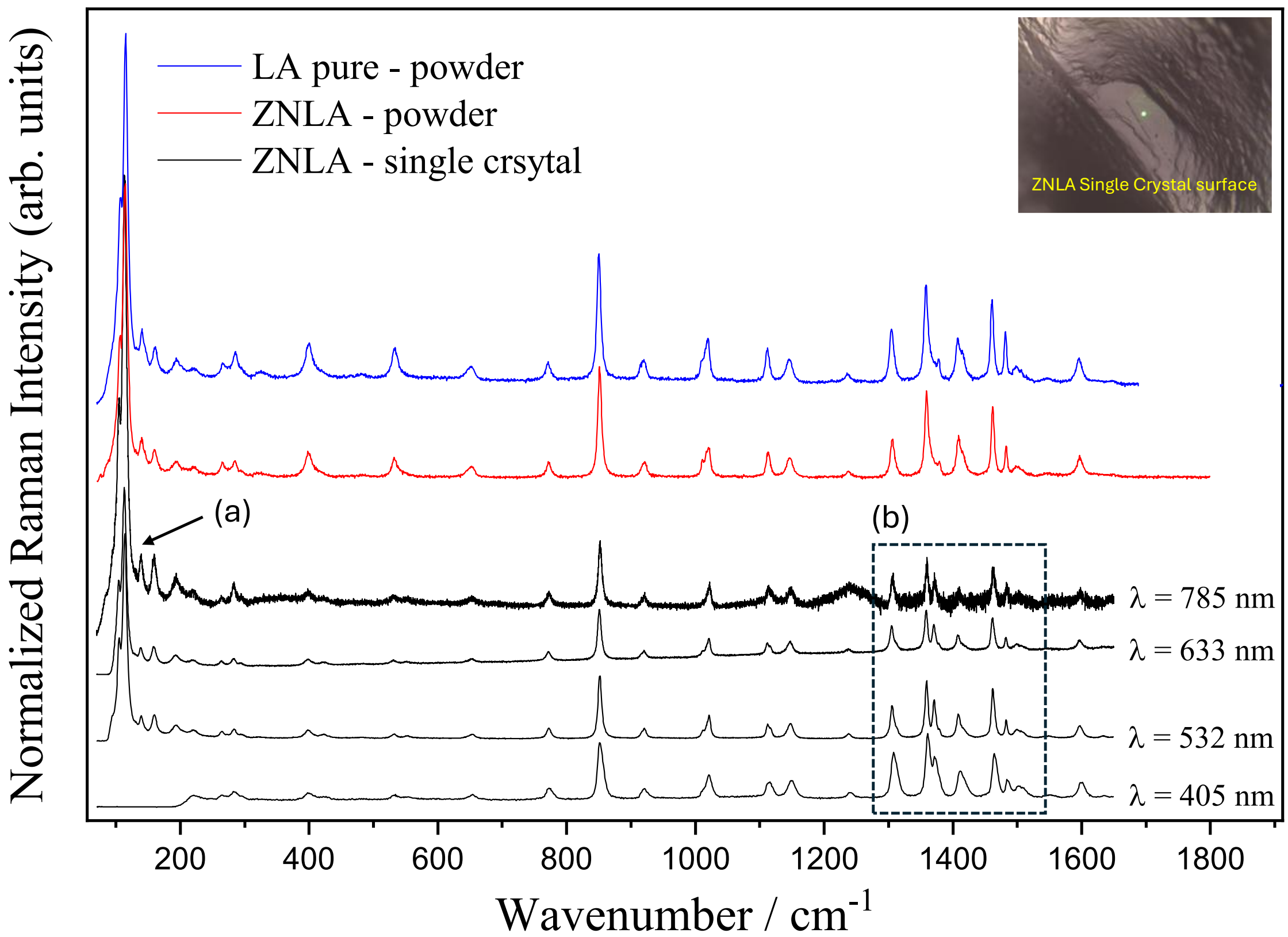

**Figure 3.** Raman spectra of Zn-doped L-alanine (ZNLA) single crystal and LA and ZNLA powder at room temperature. The normalized Raman spectra correspond to different laser excitation lines ranging from 405 to 785 nm. The curves are vertically offset for clarity. The inset shows the image of the ZNLA single-crystal surface where Raman measurements with different wavelengths were performed. The small green dot indicates the precise position of the laser spot. The arrow in (a) indicates the region of the lattice modes, and in (b), a reduction in the integrated intensity of the modes – mostly associated with $CH_3$ – can be observed.

To investigate the effect of laser excitation on phonon mode intensities in more detail, the observed phonon modes were fitted using Lorentzian functions [47]. **Figure 4 (a–b)** illustrates the peak deconvolution of the spectrum in the range of 80 to 320 $cm^{-1}$, primarily associated with lattice modes and CC´N deformations. In the 80–240 $cm^{-1}$ range, no significant changes in the overall band profiles were observed when varying the excitation wavelength from 785 to 532 nm. However, the vibrational modes at 104 and 114 $cm^{-1}$, attributed to lattice vibrations, show a pronounced increase in their normalized integrated intensities, showing variations of 37 and 15, respectively (see **Figure 6a**). Due to the edge filter limitations, modes below 200 $cm^{-1}$ were not detected with the 405 nm laser line. On the other hand, in the range of 240 to 320 $cm^{-1}$, notable broadening of the bands and subtle changes in the spectra profile were observed across the entire excitation range. The peaks observed at 264 and 285 $cm^{-1}$, associated with deformations of the carboxyl group (e.g., def. CC´N), did not exhibit appreciable variations in their integrated intensities (see **Figure 6b-c**). Particularly, the mode observed at 264 $cm^{-1}$ is not expected to exhibit any absolute variation in its intensity, as it was chosen as the normalization reference for the spectra. It is shown in **Figure 6b** for reference purposes only.

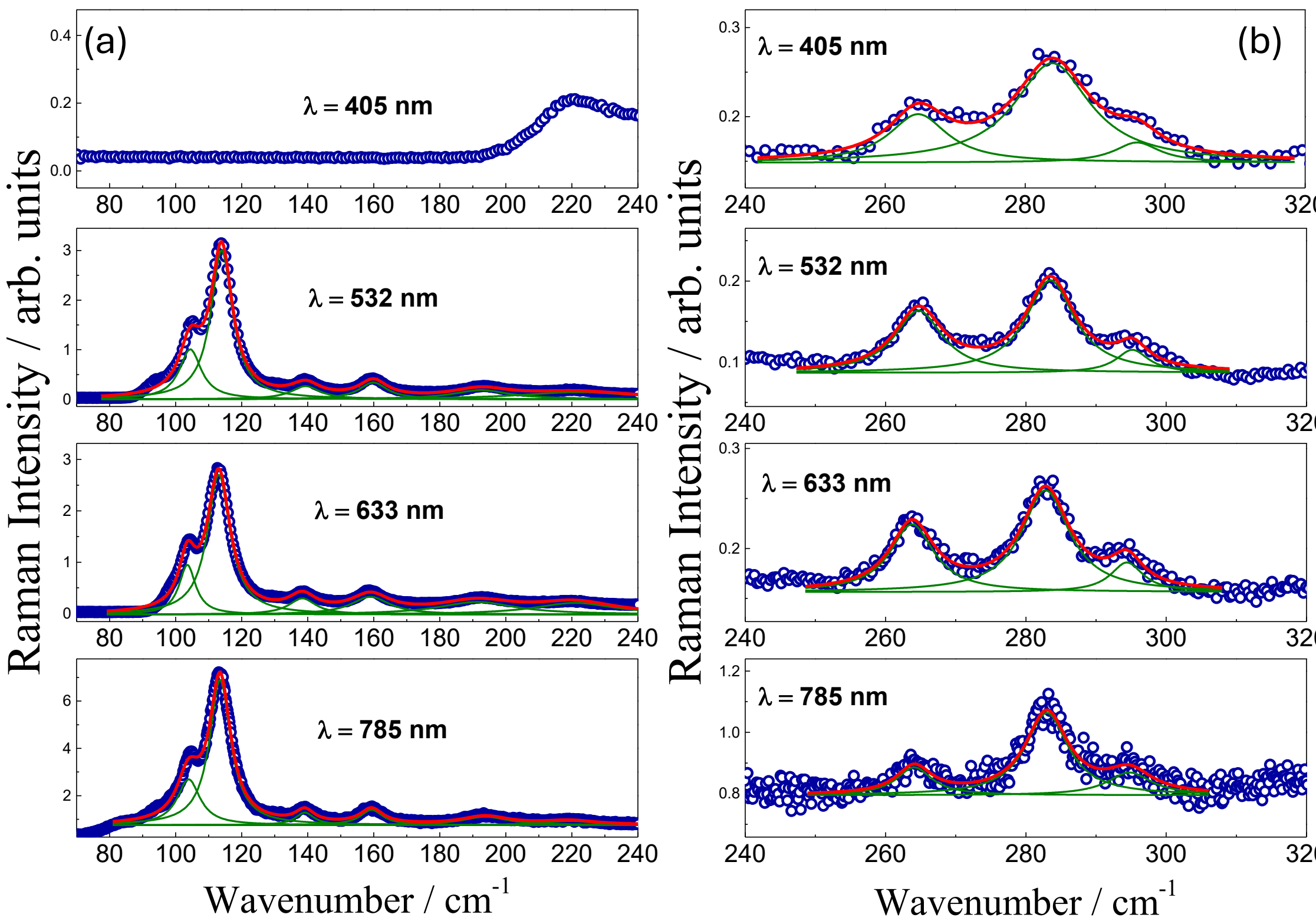

**Figure 4.** Evolution of the Raman spectra of ZNLA single crystals under different excitation wavelengths, with corresponding peak deconvolution in the regions (a) 70–240 $cm^{-1}$ and (b) 240–320 $cm^{-1}$.

**Figure 5** shows the peak deconvolution of the Raman spectrum in the range of 700 to 1400 $cm^{-1}$. Within the range of 700 to 950 $cm^{-1}$, no significant changes in the band profile were observed for excitation wavelengths from 785 to 405 nm. The Raman modes observed at approximately 772, 852, and 918 $cm^{-1}$ are attributed to $\delta(\mathrm{CO_2}), v_s(\mathrm{CCNC})$ and $v_{as}(\mathrm{CCNC})$, respectively. As shown in **Figure 6 (e-f)**, the modes at 772 and 852 $cm^{-1}$ exhibit similar behavior in their normalized intensities under different excitation wavelengths. Specifically, their intensities gradually decrease under 405, 633, and 532 nm excitation, but increase when excited at 785 nm. On the other hand, the intensities of the vibrational modes at 1361 and 1371 $cm^{-1}$, attributed to $\delta(\mathrm{CH_3})$ and $\delta_s(\mathrm{CH_3})$, respectively, decrease under excitation from 405 to 785 nm, exhibiting variations in their normalized integrated intensities of 6.34 and 7.43, respectively (see **Figure 6(g-h)**). All these pronounced variations in normalized integrated intensities can be attributed to resonance effects due to the presence of Zn impurities in L-alanine single crystal, as also previously observed for Fe and Cu-based L - alanine compounds [23,24].

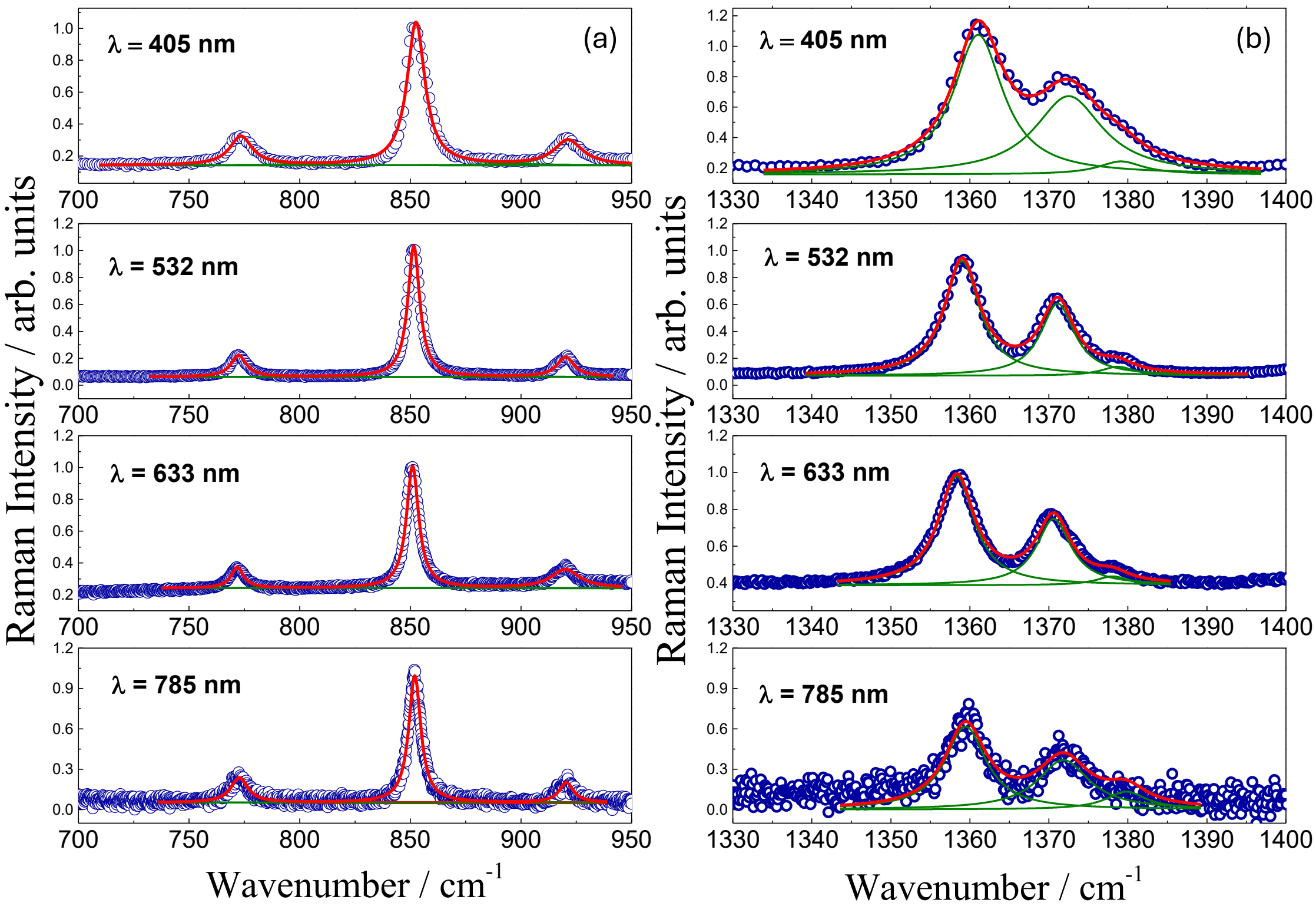

**Figure 5.** Evolution of the Raman spectra of ZNLA single crystals under different excitation wavelengths, with corresponding peak deconvolution in the regions (a) 700–950 $cm^{-1}$ and (b) 1330–1400 $cm^{-1}$.

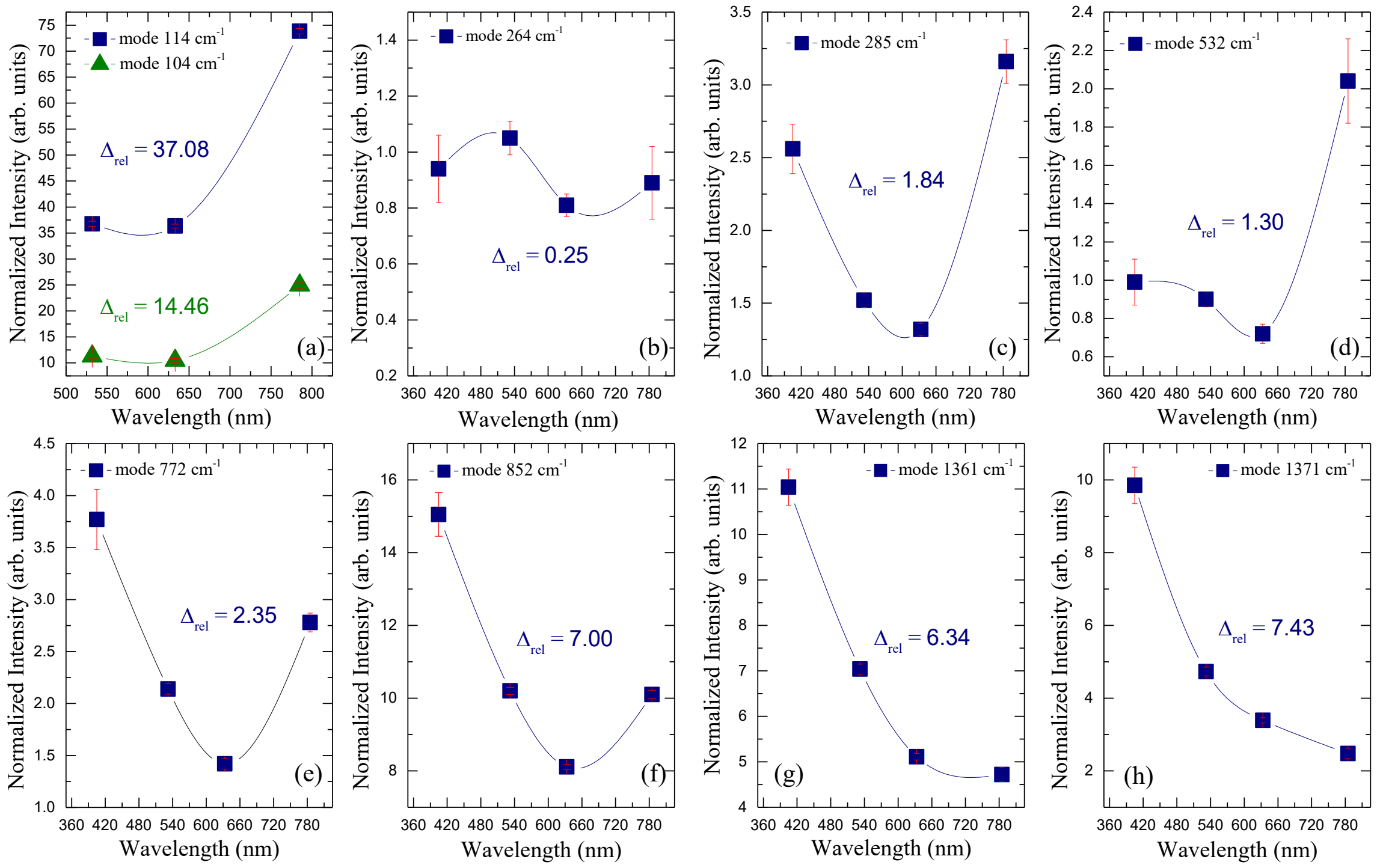

**Figure 6.** The normalized integrated intensity of the modes exhibiting resonant enhancement for the four laser excitations in the Zn-doped single crystal. The solid lines are a guide for the eyes. $\Delta_{rel}$ (arb. units) refers to the absolute variation of the normalized integrated intensity.

The Room-temperature UV-Vis absorption spectrum of the Zn-doped L-alanine crystal complements the information needed to clarify the resonance profiles observed in **Figure 6**. The resonant behavior of the selected phonons suggests that their coupling to electronic transitions occurs in a distinctly different way. Thus, in **Figure 7a**, we show the absorption spectrum of the $Zn^{2+}$-doped L-alanine crystal, where we observe an intense absorption in the ultraviolet (UV) range between 3.0 and 4.3 eV. The maximum absorption lies around 4.4 eV. Also, a weak absorption band appears between 2.0 and 3.0 eV in Zn-doped L-alanine single crystals. These results resemble those found for pure L-alanine crystal, where only low-intensity absorption is observed in the visible region [48]. This feature can be attributed to the introduction of intermediate energy levels within the band gap, arising from doping or local structural defects [44]. Furthermore, such a structure may suggest the presence of additional absorption maxima arising from intermolecular interactions, which could lead to the formation of molecular sub-bands acting as effective trapping levels within the forbidden energy gap [48]. The bandgap ($E_g$) of the sample was calculated using the Tauc plot method [49], based on the appropriate optical absorption coefficient. The bandgap energy was determined by extrapolating the straight-line portion of the curve to $(\alpha h\nu)^2 = 0$. As shown in **Figure 7b,** the obtained direct bandgap value is 3.75 eV. This bandgap value is relatively lower compared to that reported for pure L-alanine, which is typically greater than 5.0 eV [50–53]. A similar behavior was observed for both undoped and Zn-doped $Nb_2O_5$, with the doped sample exhibiting a significant reduction in bandgap compared to the undoped one [54,55]. Furthermore, DFT calculations previously reported have shown that $Li^+$, which possesses an ionic radius very similar to that of $Zn^{2+}$ (e.g., $Li^+$: ~0.76 Å; $Zn^{2+}$: ~0.74 Å), drastically reduces the bandgap of L-alanine to approximately 3.5 eV, a value quite close to that obtained experimentally for Zn-doped L-alanine (3.75 eV) [56].

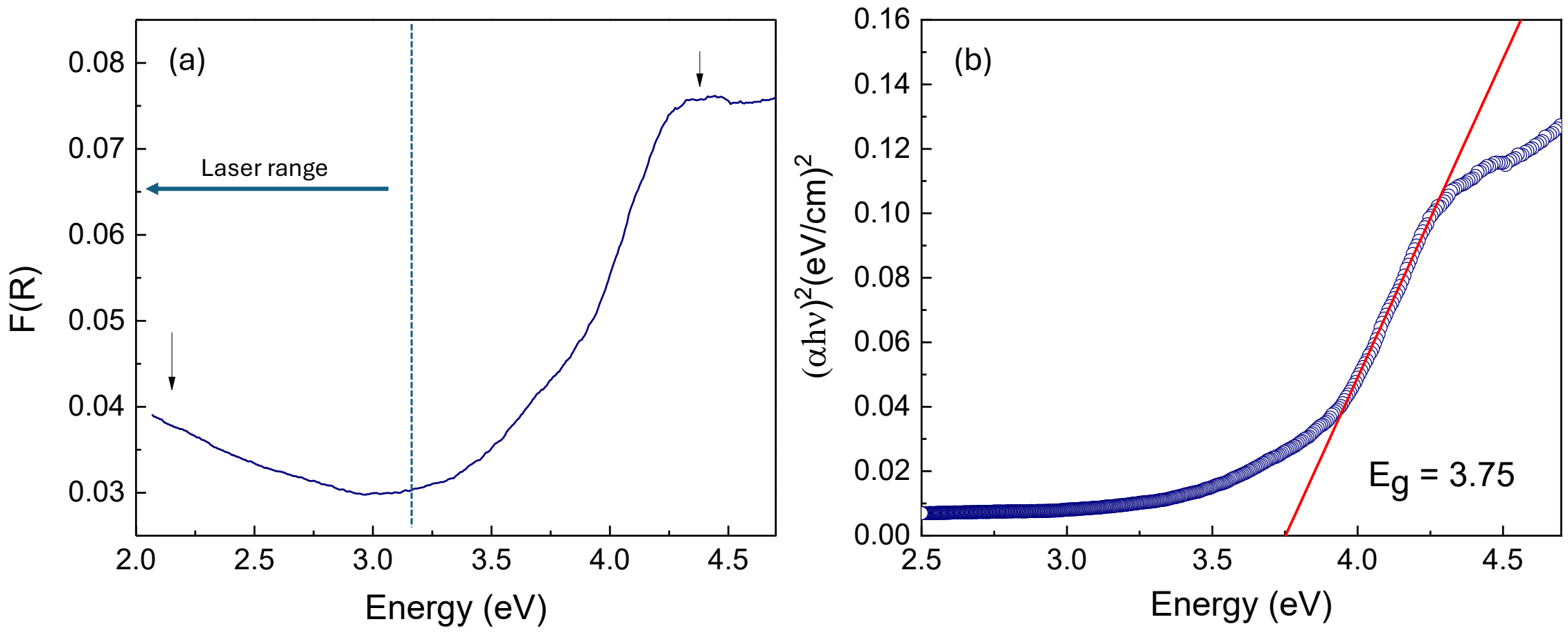

**Figure 7.** (a) The UV-vis absorption spectrum of the Zn-doped L-alanine crystal, and (b) the Tauc plot for direct bandgap ($E_g$) with the linear extrapolation adjustment. Note the presence of weak absorptions in the visible region and a strong dipole-allowed absorption in the UV region.

Therefore, the UV-Vis absorption spectrum analysis corroborates that the vibrational modes at 852, 1361, and 1371 $cm^{-1}$ exhibit resonant behavior, with their maximum enhancement occurring when the laser energy at 405 nm (3.1 eV) approaches the absorption band. Typically, the first dipole-allowed electronic transition occurs within the 3.1 - 4.7 eV range for several doped-based L-alanine systems [16,24,33,57,58]. Since the Zn-doped L-alanine single crystal exhibits a direct band gap of approximately 3.8 eV, the laser excitation energy (3.1 eV) is slightly lower than the electronic transition energy in the UV region. Thus, the enhancements in the modes observed in **Figures 6(f-h)** are just due to the tails of the resonant profiles, as the laser used does not have enough energy to reach the peaks of these profiles. This result is in perfect agreement with the observations made by E. Winkler *et al.*, where they observed that resonant behavior reaches its maximum enhancement at energies higher than the shortest wavelength used by the $Ar^{+}$ laser (2.7 eV) [24]. They have attributed the resonant behavior of the modes at 267.2, 285 and 532 $cm^{-1}$ to the bonding of impurities ($Fe^{3+}$ and $Cu^{2+}$) with the amino and carboxyl groups. However, we do not observe a larger variation in normalized intensity for ZNLA modes at 267.2, 285 and 532 $cm^{-1}$ as previously observed for the $Fe^{3+}/Cu^{2+}$-based L-alanine compound. On the other hand, this resonant effect is more prominent in the modes at 852, 1361, and 1371 $cm^{-1}$, which are associated with intramolecular vibrations of the carboxyl and methyl groups. Conversely, the lattice modes associated with hydrogen bonds (N–H···O) between the COOH group and neighboring amino groups exhibited a strong resonant enhancement under 785 nm excitation, unlike the other modes. Based on this evidence, we propose that zinc interacts with amino and carboxyl groups. **Figure 8** shows a projection on the $ac$ plane of the unit cell of L-alanine as obtained from X-ray diffraction, including the proposed position of interstitial zinc ions as suggested by Takeda *et al.* [59] for the copper ion. It is important to highlight that, as observed by E. Winkler *et al.*[24], our results indicate large contributions of the zinc dopant on L-alanine crystal lattice modes intensity, particularly at longer excitation wavelengths (e.g., 785 nm).

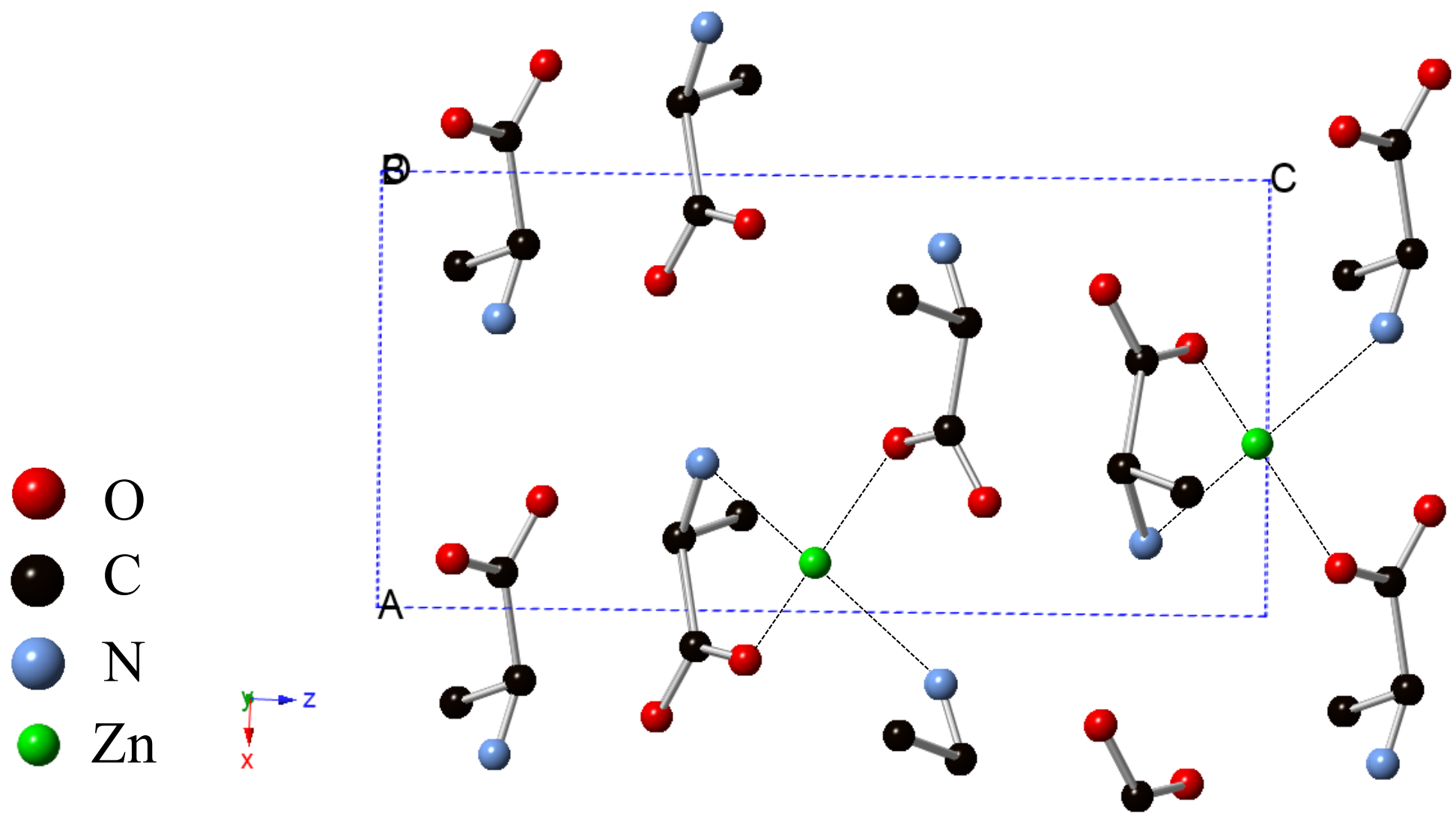

**Figure 8**. Projection of the crystal structure of L-alanine crystal on the (010) plane. The Zn(II) position is assumed to be identical to that proposed by Takeda *et al.* for Cu(II) [59]. Each Zn(II) atom is locally coordinated by two oxygen and two nitrogen atoms, forming a planar array with $D_{2h}$ symmetry (ignoring distortions) at the impurity site. To better clarify, the hydrogen atoms are omitted.

To shed light on the strong resonant effect exhibited by the 115 $cm^{-1}$ mode in the Zn-doped L-alanine (ZNLA) single crystal (see **Figure 6a**), it is worth noting that, according to classical inelastic scattering theory, the one-phonon Raman cross section depends on the derivative of the polarizability tensor ($\boldsymbol{\alpha}$) with respect to the normal mode coordinate ($q$). This cross section is mainly influenced by how $\boldsymbol{\alpha}$ ($\omega$) varies with electronic energy, leading to resonance when an optically allowed transition is near. Thus, in RRS, the intensity of the Raman line depends on the term $I \propto \left|\frac{\partial \propto(\omega)}{\partial q}\right|^2$ [60]. Thus, if interstitial $Zn^{2+}$ is slightly altering the local environment of the lattice modes (even without distorting the overall crystal cell unit), it may increase the polarizability associated with these collective vibrations, thereby enhancing the Raman scattering cross-section of these lattice modes. Thus, we observed electronic absorptions in the visible region that couple more strongly with the lattice modes, as previously reported for Fe-doped L-alanine single crystals [44]. The differences observed in absorption spectra profile in the visible region between Zn-doped and pure L-alanine single crystals, although $Zn^{2+}$ has a closed-shell $3d^{10}$ configuration, may be attributed to transitions involving defect-induced states or local electronic environments that slightly perturb the crystal field symmetry [61,62].

## 4. Conclusion

In this study, we investigated the effects of Zn impurities on the structural and vibrational properties of the L-alanine structure. Single crystals of Zn-doped L-alanine (ZNLA) were grown using the slow evaporation technique at room temperature. The X-ray diffraction (XRD) results confirm that no structural phase transition was observed for low concentrations of zinc (< 5%), with ZNLA single crystals crystallising in an orthorhombic structure belonging to space group $P2_12_12_1$. Using Raman spectroscopy with different fixed excitation laser lines (405, 532, 633, and 785 nm) we investigated the effect of $Zn^{2+}$ impurities on the intensity of Raman-active modes. Analysis of the normalized intensity of the vibrational modes revealed a selective resonant enhancement of Raman-active modes related to intramolecular vibrations of the methyl and carboxyl groups. Moreover, the vibrational mode that exhibits the greatest variation in its intensity corresponds to the lattice mode at 115 $cm^{-1}$, which is associated with intermolecular hydrogen bonding interactions between the carboxyl and amino groups of neighboring molecules (N–H···O). These findings suggest that Zinc occupies an interstitial position interacting with amino and carboxyl groups, being Zn impurities bounding to the ZNLA crystal in the same way as in Cu- and Fe-doped L-alanine crystals. The UV-Vis optical absorption measurements revealed intense absorption in the ultraviolet (UV) range between 3.0 and 4.3 eV, while a weaker absorption band was observed between 2.0 and 3.0 eV in Zn-doped L-alanine single crystals. These observed features point to subtle intermolecular interactions, possibly giving rise to additional absorption maxima, indicating the formation of molecular sub-bands acting as effective trapping levels within the forbidden energy gap. Additionally, these measurements allowed for the determination of the direct bandgap, which was found to be 3.75 eV. This approach proves to be useful for detecting interactions between transition metal ion impurities and intramolecular structures in other transition metal-doped organic amino acid systems, suggesting that most doped L-alanine compounds should be coordinated similarly to the Zn-doped L-alanine compound.

**Acknowledgements**

The authors are grateful for financial support from CNPq (Project No 431943/2016-8, Grant No. PQ-310127/2023-8), CAPES (Finance Code 001) and FAPEMA (Project No. COOPI-07771/17; BPD – 05073/21, Grant No. INFRA-02050/21) for co-funding this work.

**References**

[1] S. Sathiyanathan, M. Selvapandiyan, Enhancement of yttrium ion on the properties of L-histidine nitrate single crystal for frequency conversion applications, Materials Science- Poland 37 (2019) 615–621. https://doi.org/10.2478/MSP-2019-0071.

[2] S. Chandran, R. Paulraj, P. Ramasamy, Crystal growth, piezoelectric, non-linear optical and mechanical properties of lithium hydrogen oxalate monohydrate single crystal, AIP Conf Proc 1832 (2017). https://doi.org/10.1063/1.4980613/1020513.

[3] S. Velayutham, M. Selvapandiyan, Effect of yttrium ion on the properties of tri ethyl ammonium picrate single crystals, Heliyon 5 (2019) e02091. https://doi.org/10.1016/J.HELIYON.2019.E02091.

[4] X. Mu, K. Xu, J. Ma, T. Wang, X. Feng, Y. Zhai, F. Xuan, L. Cao, B. Teng, Organic second-order nonlinear optical crystals: materials for terahertz, Journal of Materials Science: Materials in Electronics 35 (2024) 1–16. https://doi.org/10.1007/S10854-024-12069-7/METRICS.

[5] S.Y. Zhang, X. Shu, Y. Zeng, Q.Y. Liu, Z.Y. Du, C.T. He, W.X. Zhang, X.M. Chen, Molecule-based nonlinear optical switch with highly tunable on-off temperature using a dual solid solution approach, Nat Commun 11 (2020) 1–8. https://doi.org/10.1038/S41467-020-15518-Z;TECHMETA=128,145;SUBJMETA=263,298,638,639,911,915,921;KWRD=COORDINATION+CHEMISTRY,METAL.

[6] K. Kamatchi, S. Sriram, G. Vinitha, Exploring an efficient nonlinear optical single-crystal L-Proline potassium pentaborate octahydrate (LPPPB) for optical limiting applications, Journal of Materials Science: Materials in Electronics 33 (2022) 4694–4703. https://doi.org/10.1007/S10854-021-07657-W/FIGURES/9.

[7] Y. Zhang, X. Bai, J. Arias Muñoz, Y. Dai, S. Das, Y. Wang, Z. Sun, Coherent modulation of chiral nonlinear optics with crystal symmetry, Light Sci Appl 11 (2022) 1–7. https://doi.org/10.1038/S41377-022-00915-4;TECHMETA=128,133,140;SUBJMETA=1130,2799,385,400,624,639,766;KWRD=NONLINEAR+OPTICS,PHOTONIC+DEVICES.

[8] P. Gandhimathi, Promising LVHCL material for nonlinear optical devices and pyroelectric sensors, Journal of Optics (India) (2024) 1–6. https://doi.org/10.1007/S12596-024-02292-0/TABLES/4.

[9] H. Xu, J. Li, T. Luo, W. Zhang, H. Jin, Y. Gao, T. He, Nonlinear Optical Properties of Cs2AgIn1-xBixCl6 Single Crystals, Journal of Physical Chemistry Letters 15 (2024) 5848–5853. https://doi.org/10.1021/ACS.JPCLETT.4C00986/ASSET/IMAGES/LARGE/JZ4C00986_0006.JPEG.

[10] Z. Yan, J. Fan, S. Pan, M. Zhang, Recent advances in rational structure design for nonlinear optical crystals: leveraging advantageous templates, Chem Soc Rev 53 (2024) 6568–6599. https://doi.org/10.1039/D3CS01136D.

[11] R.W. Boyd, Nonlinear Optics, Third Edition, Elsevier, 2008.

[12] A. Rakini, K. Rajarajan, M. Neela, B. Premalatha, S. Surya, Growth, characterization and second harmonic generation NLO activity of semi-organic crystal: l-arginine picrate crystal doped with nickel chloride, Journal of Materials Science: Materials in Electronics 35 (2024) 1–12. https://doi.org/10.1007/S10854-024-12445-3/METRICS.

[13] R. Takeda, A. Kawamura, A. Kawashima, T. Sato, H. Moriwaki, K. Izawa, H. Abe, V.A. Soloshonok, Second-order asymmetric transformation and its application for the practical synthesis of α-amino acids, Org Biomol Chem 16 (2018) 4968–4972. https://doi.org/10.1039/C8OB00963E.

[14] K. Vaiyapuri, T. Subramani, A.K. Rajamani, M.L. Thangavel, S.K. Ganesan, S. Palanisamy, K. Malaivelusamy, Organometallic L-Alanine Cadmium Iodide Crystals for Optical Device Fabrication, Journal of Electronic Science and Technology 20 (2022) 345–355. https://doi.org/10.1016/j.jnlest.2022.100178.

[15] D.P. Karothu, G. Dushaq, E. Ahmed, L. Catalano, S. Polavaram, R. Ferreira, L. Li, S. Mohamed, M. Rasras, P. Naumov, Mechanically robust amino acid crystals as fiber-optic transducers and wide bandpass filters for optical communication in the near-infrared, Nat Commun 12 (2021) 1–8. https://doi.org/10.1038/s41467-021-21324-y.

[16] B.S. Benila, K.C. Bright, S.M. Delphine, R. Shabu, Optical, thermal and magnetic studies of pure and cobalt chloride doped L-alanine cadmium chloride, J Magn Magn Mater 426 (2017) 390–395. https://doi.org/10.1016/j.jmmm.2016.11.120.

[17] P.M. Wankhade, G.G. Muley, Growth, morphology, optical, thermal, mechanical and electrical studies of a cesium chloride doped L-alanine single crystal, Chinese Journal of Physics 55 (2017) 2181–2191. https://doi.org/10.1016/j.cjph.2017.06.015.

[18] D.A. Fentaw, M.E. Peter, T. Abza, Synthesis and characterization of lanthanum chloride doped L-alanine maleate single crystals, J Cryst Growth 522 (2019) 1–4. https://doi.org/10.1016/j.jcrysgro.2019.06.002.

[19] N. Suresh, M. Selvapandiyan, F. Mohammad, S. Sagadevan, Influence of bismuth nitrate doping towards the characteristics of L-Alanine nonlinear optical crystals, Chinese Journal of Physics 67 (2020) 349–359. https://doi.org/10.1016/j.cjph.2020.07.014.

[20] G. Krishnamoorthi, R. Uvarani, Growth and characterization of pure and metals-doped organic nonlinear optical single crystal: l-alanine alaninium nitrate (LAAN), Journal of Materials Science: Materials in Electronics 32 (2021) 3979–3988. https://doi.org/10.1007/s10854-020-05140-6.

[21] N. Suresh, M. Selvapandiyan, Influence of Zirconium Nitrate doping on the properties of l-Alanine crystal for nonlinear optical applications, Journal of Materials Science: Materials in Electronics 31 (2020) 16737–16745. https://doi.org/10.1007/S10854-020-04229-2/TABLES/2.

[22] G.S. Kumar, J.M.S. Gnanaraj, V. Kathiravan, P. Karuppasamy, M.S. Pandian, P. Ramasamy, Growth of potassium iodide-doped L-alanine nonlinear optical single crystals: investigation of physico-chemical properties, Journal of Materials Science: Materials in Electronics 33 (2022) 26764–26774. https://doi.org/10.1007/S10854-022-09342-Y/TABLES/3.

[23] E. Winkler, A. Fainstein, P. Etchegoin, C. Fainstein, Fe impurities in l-alanine: an epr, luminescence, and raman study, Phys Rev B Condens Matter Mater Phys 59 (1999) 1255–1262. https://doi.org/10.1103/PhysRevB.59.1255.

[24] E. Winkler, P. Etchegoin, A. Fainstein, C. Fainstein, Resonant Raman scattering and optical transmission studies of Cu(II) and Fe(III) impurities in crystalline L-alanine, Phys Rev B Condens Matter Mater Phys 61 (2000) 15756–15761. https://doi.org/10.1103/PhysRevB.61.15756.

[25] D. Sankar, P. Praveen Kumar, J. Madhavan, Influence of metal dopants (Cu and Mg) on the thermal, mechanical and optical properties of l-alanine acetate single crystals, Physica B Condens Matter 405 (2010) 1233–1238. https://doi.org/10.1016/j.physb.2009.11.048.

[26] A.L.O. Cavaignac, R.J.C. Lima, P.F. Façanha Filho, A.J.D. Moreno, P.T.C. Freire, High-temperature Raman study of L-alanine, L-threonine and taurine crystals related to thermal decomposition, Physica B Condens Matter 484 (2016) 22–26. https://doi.org/10.1016/j.physb.2015.12.044.

[27] C.H. Wang, R.D. Storms, Temperature-Dependent Raman Study and Molecular Motion in l-Alanine Single Crystal, J Chem Phys 55 (1971) 3291–3299. https://doi.org/10.1063/1.1676579.

[28] A.M.R. Teixeira, P.T.C. Freire, A.J.D. Moreno, J.M. Sasaki, A.P. Ayala, J. Mendes Filho, F.E.A. Melo, High-pressure Raman study of L-alanine crystal, Solid State Commun 116 (2000) 405–409. https://doi.org/10.1016/S0038-1098(00)00342-2.

[29] B. Suresh Kumar, M.R. Sudarsana Kumar, K. Rajendra Babu, Growth and characterization of pure and lithium doped L-alanine single crystals for NLO devices, Crystal Research and Technology 43 (2008) 745–750. https://doi.org/10.1002/crat.200711109.

[30] N. Suresh, M. Selvapandiyan, Influence of Zirconium Nitrate doping on the properties of l-Alanine crystal for nonlinear optical applications, Journal of Materials Science: Materials in Electronics 31 (2020) 16737–16745. https://doi.org/10.1007/s10854-020-04229-2.

[31] M.R. Hareeshkumar, M.R. Jagadeesh, G.J. Shankaramurthy, B.M. Prasanna, Growth, Nonlinear Optical, Electrical, Mechanical and Dielectric Properties of Zinc Sulphate Doped L-Alanine Single Crystal for Optoelectronic Applications, IOP Conf Ser Mater Sci Eng 1166 (2021) 012035. https://doi.org/10.1088/1757-899x/1166/1/012035.

[32] A.X. Trautwein, Bioinorganic chemistry : transition metals in biology and their coordination chemistry, 1977.

[33] D. Jini, M. Aravind, L. Jothi Nirmal, S. Ajitha, Structural, optical, and biological properties of L-alanine single crystal by slow evaporation method, Mater Today Proc 43 (2020) 2032–2035. https://doi.org/10.1016/j.matpr.2020.11.736.

[34] A.C. Larson, R.B. Von Dreele, General structure analysis system (GSAS), Los Alamos Natl. Lab. Rep. LAUR (2000) 86–748. https://cir.nii.ac.jp/crid/1573668926168600320.

[35] M. Wojdyr, Fityk: a general-purpose peak fitting program, J Appl Crystallogr 43 (2010) 1126–1128. https://doi.org/10.1107/S0021889810030499.

[36] D. Jaikumar, S. Kalainathan, G. Bhagavannarayana, Structural, spectral, thermal, dielectric, mechanical and optical properties of urea l-alanine acetate single crystals, Physica B Condens Matter 405 (2010) 2394–2400. https://doi.org/10.1016/J.PHYSB.2010.02.053.

[37] R.D. Shannon, Revised effective ionic radii and systematic studies of interatomic distances in halides and chalcogenides, Acta Crystallographica Section A 32 (1976) 751–767. https://doi.org/10.1107/S0567739476001551.

[38] N. Vijayan, S. Rajasekaran, G. Bhagavannarayana, R. Ramesh Babu, R. Gopalakrishnan, M. Palanichamy, P. Ramasamy, Growth and characterization of nonlinear optical amino acid single crystal: L-Alanine, Cryst Growth Des 6 (2006) 2441–2445. https://doi.org/10.1021/CG049594Y/ASSET/IMAGES/MEDIUM/CG049594YN00001.GIF.

[39] C.H. Wang, R.D. Storms, Temperature-Dependent Raman Study and Molecular Motion in l-Alanine Single Crystal, J Chem Phys 55 (1971) 3291–3299. https://doi.org/10.1063/1.1676579.

[40] J. Han, Y.S. Chi, Vibrational and electronic spectroscopic characterizations of amino acid-metal complexes, J Appl Biol Chem 53 (2010) 821–825. https://doi.org/10.3839/JKSABC.2010.124/METRICS.

[41] K. Machida, A. Kagayama, Y. Saito, Polarized Raman spectra and intermolecular potential of DL-alanine crystal, Journal of Raman Spectroscopy 7 (1978) 188–193. https://doi.org/10.1002/jrs.1250070405.

[42] C.H. Wang, R.D. Storms, Raman Study of Hydrogen Bonding and Long-Wavelength Lattice Modes in an L-Alanine Single Crystal, J Chem Phys 55 (1971) 5110–5119. https://doi.org/10.1063/1.1675629.

[43] J. Bandekar, L. Genzel, F. Kremer, L. Santo, The temperature-dependence of the far-infrared spectra of l-alanine, Spectrochim Acta A 39 (1983) 357–366. https://doi.org/10.1016/0584-8539(83)80010-5.

[44] E. Winkler, P. Etchegoin, A. Fainstein, C. Fainstein, Resonant Raman scattering and optical transmission studies of Cu(II) and Fe(III) impurities in crystalline L-alanine, Phys Rev B Condens Matter Mater Phys 61 (2000) 15756–15761. https://doi.org/10.1103/PhysRevB.61.15756.

[45] C.H. Wang, R.D. Storms, Temperature-Dependent Raman Study and Molecular Motion in l-Alanine Single Crystal, J Chem Phys 55 (1971) 3291–3299. https://doi.org/10.1063/1.1676579.

[46] E. Winkler, A. Fainstein, P. Etchegoin, C. Fainstein, Fe impurities in l-alanine: an epr, luminescence, and raman study, Phys Rev B Condens Matter Mater Phys 59 (1999) 1255–1262. https://doi.org/10.1103/PhysRevB.59.1255.

[47] M. Wojdyr, Fityk: A general-purpose peak fitting program, J Appl Crystallogr 43 (2010) 1126–1128. https://doi.org/10.1107/S0021889810030499.

[48] A. Wojciechowski, K. Ozga, A.H. Reshak, R. Miedzinski, I. V. Kityk, J. Berdowski, Z. Tylczyński, Photoinduced effects in l-alanine crystals, Mater Lett 64 (2010) 1957–1959. https://doi.org/10.1016/J.MATLET.2010.06.034.

[49] Y. Vignollet, J.C. Maire, OPTICAL PROPERTIES AND ELECTRONIC STRUCTURE OF AMORPHOUS Ge AND Si, Chemical Communications 3 (1968) 1187.

[50] P.M. Wankhade, G.G. Muley, Growth, morphology, optical, thermal, mechanical and electrical studies of a cesium chloride doped L-alanine single crystal, Chinese Journal of Physics 55 (2017) 2181–2191. https://doi.org/10.1016/j.cjph.2017.06.015.

[51] N. Suresh, M. Selvapandiyan, F. Mohammad, S. Sagadevan, Influence of bismuth nitrate doping towards the characteristics of L-Alanine nonlinear optical crystals, Chinese Journal of Physics 67 (2020) 349–359. https://doi.org/10.1016/j.cjph.2020.07.014.

[52] N. Suresh, M. Selvapandiyan, Influence of Zirconium Nitrate doping on the properties of l-Alanine crystal for nonlinear optical applications, Journal of Materials Science: Materials in Electronics 31 (2020) 16737–16745. https://doi.org/10.1007/S10854-020-04229-2/TABLES/2.

[53] N. Suresh, M. Selvapandiyan, P. Sakthivel, K. Loganathan, Structural, optical, thermal, and magnetic properties of strontium nitrate doped L-Alanine crystal, Optik (Stuttg) 221 (2020). https://doi.org/10.1016/j.ijleo.2020.165336.

[54] A. Thirugnanam, K. Rajesh, R.S. Selvaraj, P. Praveen Kumar, Growth and Characterization of Pure and Zinc Doped L-Alanine Alaninium Nitrate Single Crystals, (n.d.).

[55] J.A. Oliveira, M.O. Reis, M.S. Pires, L.A.M. Ruotolo, T.C. Ramalho, C.R. Oliveira, L.C.T. Lacerda, F.G.E. Nogueira, Zn-doped Nb2O5 photocatalysts driven by visible-light: An experimental and theoretical study, Mater Chem Phys 228 (2019) 160–167. https://doi.org/10.1016/J.MATCHEMPHYS.2019.02.062.

[56] H.A. Rahnamaye Aliabad, Z. Mojarradi, B.G. Yalcin, DFT studies for optoelectronic properties of pure l-alanine and doped with Li, Journal of Materials Science: Materials in Electronics 27 (2016) 4887–4897. https://doi.org/10.1007/S10854-016-4372-1/FIGURES/7.

[57] N. Suresh, M. Selvapandiyan, P. Sakthivel, K. Loganathan, Structural, optical, thermal, and magnetic properties of strontium nitrate doped L-Alanine crystal, Optik (Stuttg) 221 (2020). https://doi.org/10.1016/j.ijleo.2020.165336.

[58] P.M. Wankhade, G.G. Muley, Growth, morphology, optical, thermal, mechanical and electrical studies of a cesium chloride doped L-alanine single crystal, Chinese

Journal of Physics 55 (2017) 2181–2191. https://doi.org/10.1016/j.cjph.2017.06.015.

[59] K. Takeda, Y. Arata, S. Fujiwara, ESR study of single crystals of copper(II)-doped L-alanine, J Chem Phys 53 (1970) 854. https://doi.org/10.1063/1.1674080.

[60] P. Jorgensen, J. Simons, Geometrical Derivatives of Energy Surfaces and Molecular Properties, Springer Netherlands, 1986. https://doi.org/10.1007/978-94-009-4584-5.

[61] J. Zhu, T. Liu, W. Hong, H. Hu, L. Zhao, First-principles investigation on the impact of interstitial Mg in KDP crystals on laser-induced damage threshold, J Solid State Chem 348 (2025) 125399. https://doi.org/10.1016/J.JSSC.2025.125399.

[62] C.E. Patrick, Y. Huang, L.H. Lewis, J.B. Staunton, Theory of Defect-Induced Crystal Field Perturbations in Rare-Earth Magnets, Phys Rev Lett 132 (2024) 056703. https://doi.org/10.1103/PHYSREVLETT.132.056703/SUPPLEMENTAL.PDF.